# Shesop Healthcare: Stress and Influenza Classification Using Support Vector Machine Kernel

Andrien Ivander Wijaya[*1], Ary S. Prihatmanto[*2], Rifki Wijaya[*3]

[*]School of Electrical Engineering and Informatics

Institut Teknologi Bandung, Jalan Ganesha 10, Bandung 40132, Indonesia

[1]`andrien@outlook.com`
[2]`asetijadi@lskk.ee.itb.ac.id`
[3]`rifkiwijaya@gmail.com`

*Abstract*—Shesop is an integrated system to make human lives more easily and to help people in terms of healthcare. Stress and influenza classification is a part of Shesop's application for a healthcare devices such as smartwatch, polar and fitbit. The main objective of this paper is to classify a new data and inform whether you are stress, depressed, caught by influenza or not. We will use the heart rate data taken for months in Bandung, analyze the data and find the Heart rate variance that constantly related with the stress and flu level. After we found the variable, we will use the variable as an input to the support vector machine learning. We will use the lagrangian and kernel technique to transform 2D data into 3D data so we can use the linear classification in 3D space. In the end, we could use the machine learning's result to classify new data and get the final result immediately: stress or not, influenza or not.

## I. Introduction

Healthcare is a core of human's life. Being healthy is one of the main objective of life ever since BC. According to Oxford's English dictionary, health is the state of being free from illness or injury and that's why for more than a thousand years, human always constantly seek the cure for all disease that spread among the world.

Health is maintained and improved by the lifestyles, social, happiness and even willingness to live. University of Illinois's research said that people with happy and lively life tend to have a higher lifespan. A long time ago, people only maintain the health using traditional method. During the time being, people invent new instruments for healthcare and it could be anything from the birth-chair, artificial body parts, even doing countless thing to improve the health standard. The traditional method still being used until now; but it's not the same. Nowadays, healthcare is not only about the doctor and a medicine but also covers advancement and application in terms of health science.

Since electrical devices, computer, artificial intelligence, big data and cognitive was found and invented, healthcare comes to the point where we can do something that can't be done before: a simple way to know your current condition and to predict your future health condition in touch of a hand of everyone, everywhere as simple as using a smartphone or a smartwatch. We can find a smartwatch and devices in the market equipped with a heartrate sensor and we can monitor our own heartrate, but it's just a numbers if we can't make a proper use of the heart rate's record.

The most popular tool to analyze heart rate's trends is by using Heart Rate Variance (HRV). HRV includes time-domain parameters, frequency-domain parameters and nonlinear parameters. Many paper, research and study believes that each parameter has their own function and effect to any physical and mentality changes over our body and could be linked to a certain condition such as sleep, running, disease and even cancer.

By using cognitive machine learning, this paper's study and research use the time-domain parameters to classify whether a person is caught by influenza or not and whether a person is depressed or not from the heart rate data from a daily constant activity. Hopefully, it could help people who couldn't or doesn't want to reach a proper health support and make them more aware with their health condition and reach the proper medical doctor or medical services; also help people to regularly monitor their condition anywhere and anytime.

## II. Background

### A. Heart Rate

Heart rate is a pulse beats (Ventricle Contraction) per certain time. Usually, we used 60 seconds to measure heart rate. Heart rate is a dynamic variable which changes over time. HR use bpm which can be measured by count the total beats in 1 minute period. Heart beats can be found in the chest, arm, neck, feet and many other places. Factors that change the heart rate:

| Activity | What you do: sleeping, running, swimming, etc |
|---|---|
| Temperature and density | High temperature and density will lead to higher heart rate. Temperature and density will not change the heart rate significantly. As the temperature increases from $16^0$C to $24^0$C degrees, a runner's heart rate at a given speed increases by about 2 to 4 beats per minute. When the temperature increases from $24^0$C to $32^0$C, you can expect your heart rate running at a given speed to increase by approximately 10 beats per minute. High humidity magnifies the effect of high temperatures on heart rate. |
| Age | Age will increase resting heart rate but decrease the maximum activity heart rate. |
| Morning VS evening | Heart rate tends to be lower in the morning. The difference in heart rate between running in the morning and afternoon is typically about 5 to 6 bpm |
| Body position | Body position changes the Heart Rate, the more you relaxed, it will lead to lower heart rate. The changes in heart rate by body condition could be very vary depends on the position. |
| Emotion | Happy, sad, angry, etc; each will lead to a different heart rate. Anger and fear could increase heart rate up to 8 bpm. Sadness could increase heart rate up to 7 bpm. Happiness could increase the heart rate up to 6 bpm while surprise and disgust could increase the heart rate up to 2 bpm |
| Dehydration and severe hunger | Dehydration and hunger could increase heart rate. A 1992 study by S. J. Montain and Ed Coyle, PhD, found that heart rate heart rate increases approximately 7 beats per minute for each 1% loss in bodyweight due to dehydration |
| Amount of sleep time | Lack of sleep will increase heart rate up to 1 bpm |
| Obese | Obese usually leads to the abnormal heart rate. Heart rate could increase 1 to 4 bpm depend on the Body Mass Index. |
| Drugs, caffeine, cigar, medicine | Cigar and caffeine could increase heartrate up to 14 bpm while certain medicine and drugs could increase and decrease heart rate into any numbers |
| Athlete body and heart | Being athletic and fit can make heart rate down so much that for example: Miguel Indurain, five-time winner of the Tour de France, reported a resting heart rate of only 28 bpm due to stronger and bigger heart muscle that could pump blood more than normal person did in one beat.. |
| After eating | After eating something, our heart rate will increased due to heart that will help body aid the digestion. More blood is directed toward the gastrointestinal tract to process the food. In some cases, resting heart rate could increase into more than 100 after eating. |

Table 1: Factor that change heart rate

## B. IBI, RR, NN and ectopic beats

We can extract IBI from RR interval (often called normal to normal NN). RR interval is the time difference between two R peaks in ECG signals. Ectopic beats is the abnormal heart beats that lead a sudden spike to ECG graph. If we ignore ectopic beats, we can assume that RRI equals to IBI.

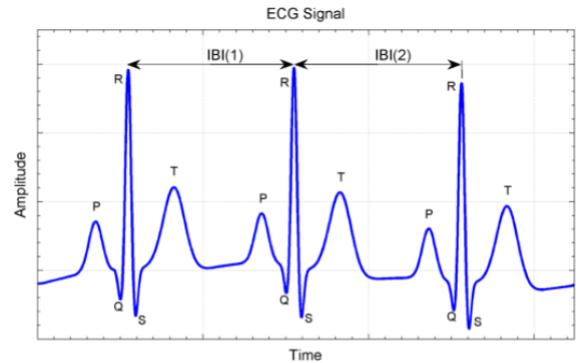

Image 1: ECG signal and PQRST

To count IBI, we can use formula:

$$IBI(n) = beat(n+1) - beat(n), \text{ for } 1 \leq n \leq N-1 \quad (1)$$

To extract IBI from HR:

$$IBI(n) = 1 / HR(n), \text{ for } 1 \leq n \leq N-1 \quad (2)$$

(HR in bps, ibi in s)

$$IBI(n) = 60 * 1000 / HR(n), \text{ for } 1 \leq n \leq N-1 \quad (3)$$

(HR in bpm, ibi in ms)

## C. Heart Rate Variance

Heart rate variance includes but not limited to:

| Name | Description | Units |
|---|---|---|
| Mean HR | Mean of HR recorded in a steady tempo | bpm |
| Mean RR | Mean of RR intervals recorded in a steady tempo | ms |
| RMSSD | Square root of the mean squared differences between successive RR intervals | Ms |
| SDevHR | Standard deviation of HR | bpm |
| SDevNN | Standard deviation of NN/RR, usually recorded is 24 hours | ms |
| SDANN | Standard deviation of NN/RR, usually recorded is 5 minutes | ms |
| α1 | Short term non-linear fluctuations | |
| α2 | Long term non-linear fluctuations | |
| SD1 | Short term poin-care variability standard deviation | Ms |
| SD2 | Long term poin-care variability standard deviation | Ms |
| ApEn(0.2) | Approximate entropy with tolerance value 0.2 | |
| LFHF | Ratio between Low Frequency and | |

| ratio | High Frequency band power |  |
| NNx | Number of successive RR interval difference that greater than x ms |  |
| pNNx | NNx divided by RRI difference count |  |

Table 2: Heart Rate Variance

## D. Support Vector Machine

Support Vector Machines (SVM) is one of the machine learning using geometry and vector helps to learn and classify things. SVM have recently gained prominence in the field of machine learning and pattern classification. Classification is achieved by realizing a linear or non-linear separation surface in the input space. The aim of a Support Vector Machine is to devise a computationally efficient way of learning good separating hyperplanes in a high dimensional feature space. In the following the construction of such a hyperplane is described using the Maximum Margin Classifier as an example of a linear machine.

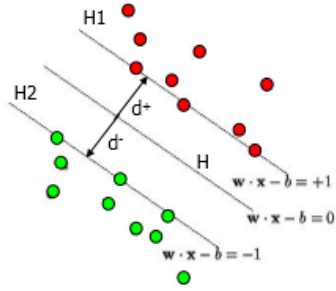

Image 2: SVM linear margin

The classifier of a Support Vector Machine can be used in a modular manner (as the kernel function) and therefore, depending on the purpose, domain, and the separable of the feature space different learners are used. For data that can be mapped using linear 2D dimension or polar dimension, we could extract the hyper-dimension by maximize the margin.

**Margin = $d^- + d^+ = 2d^-$, since $d^- = d^+$**   (4)

To maximize the margin, we need to find the distance between H1 and H2 and its:

$d^- = |w \cdot x + b|/||w|| = 1/||w||$   (5)

so, **Margin = $2/||w||$**   (6)

If the data is non-linear, we couldn't use the margin minimizer anymore but we can use some of other tools such as Lagrangian and Kernel.

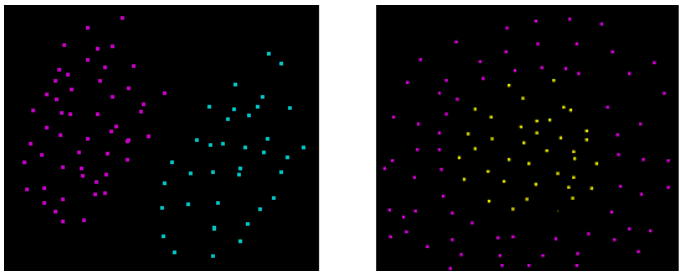

Image 3: Linear VS non-linear data

## E. Lagrangian and Gaussian Kernel

In the SVM problem, the Lagrangian equation:

$$L_P \equiv \tfrac{1}{2}\|\mathbf{w}\|^2 - \sum_{i=1}^{l} \alpha_i y_i (\mathbf{x}_i \cdot \mathbf{w} + b) + \sum_{i=1}^{l} \alpha_i$$
$$\alpha_i \geq 0, \forall i$$   (7)

Using derivatives = 0:

$$\mathbf{w} = \sum_{i=1}^{l} \alpha_i y_i \mathbf{x}_i, \sum_{i=1}^{l} \alpha_i y_i = 0$$   (8)

If we substitute the **w** into the equation,

$$\max_{\alpha \geq 0} \theta(\alpha) = \sum_i \alpha_i - \frac{1}{2} \sum_i \sum_j \alpha_i \alpha_j y_i y_j x_i^T x_j$$   (9)

With subject to

$$\sum_i \alpha_i y_i = 0$$

In the end, we want to classify the data using:

$$\omega^T x_{\text{new}} + b$$   (10)

With

$$b = y_i - \omega^T x_i$$

The result will tell the class: the result is negative or positive.

However, count the $x_i^T x_j$ is in efficient and we need something to make our counting more easily: Gaussian kernel. Technically, we could use any kernel beside of Gaussian such as Sigmoid, Polynomial, etc. By substituting $x_i^T x_j$ into a kernel function and set σ = 1, we could

$$K(x,y) = e^{\tfrac{1}{2}\|x-y\|^2}$$   (11)

Leads to:

**Max $L_p = \sum \alpha_i - \tfrac{1}{2}\sum \alpha_i \alpha_j K(x_i, x_j)$**   (12)

Lagrangian and kernel technique will mapped 2 dimensional space into a 3 dimensional space. After mapping the data, linear separation now can be done in terms of 3D space.

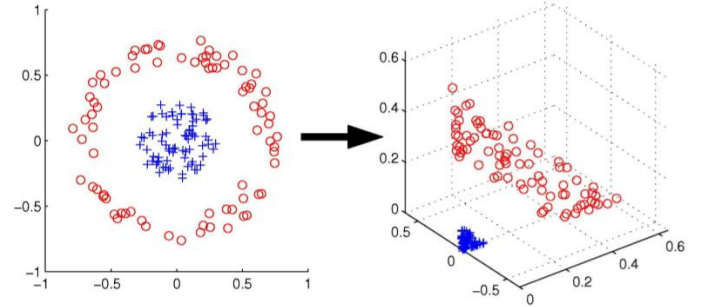

Image 4: 2D to 3D mapping

## III. DATA RECORDING

Data recording will be restricted by:

| Time | Morning, 7 AM – 8.30 AM |
|---|---|
| Temperature | Normal room temperature ($23^0C$ – $29^0C$) |
| Gender | Any |
| Age | Any |
| Activity | 20 minutes of treadmill with: <br> -10 minutes of 2km/hour walking <br> -10 minutes of 4km/hour fast walking |
| Hardward used | -Polar H7 <br> -Any Android Phone that support bluetooth low energy (Bluetooth v4) |
| Software used | Self-Developed software (Shesop Healthcare) |
| Before activity | -Recording of body temperature, sistole and diastole <br> -Sleep taken last night (in hours) <br> -Record the data: on scale 1 to 10, how much is your influenza or cough level? <br> -Record the data: on scale 1 to 10, how much is your stress or depression level? |

Table 3: Data recording scope

After 2 months of data recording, processing analyzing the heart rate variance, there are some variable that match the trend for influenza and stress level. For the stress level:

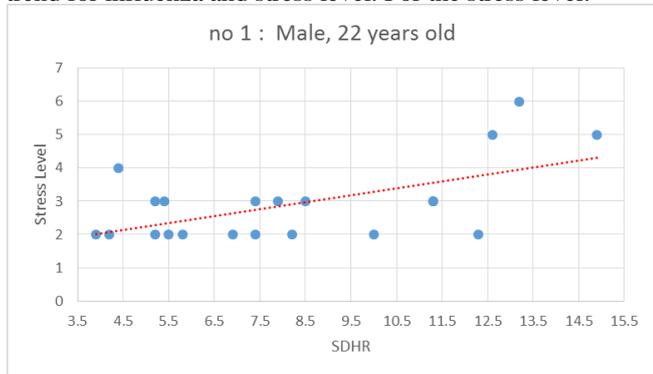

Image 5: SDHR vs Stress Level #1

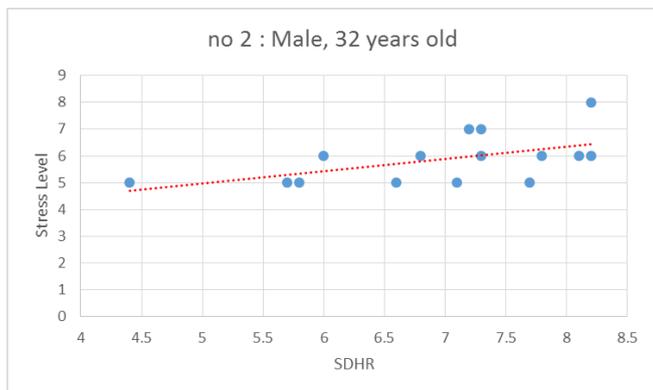

Image 6: SDHR vs Stress Level #2

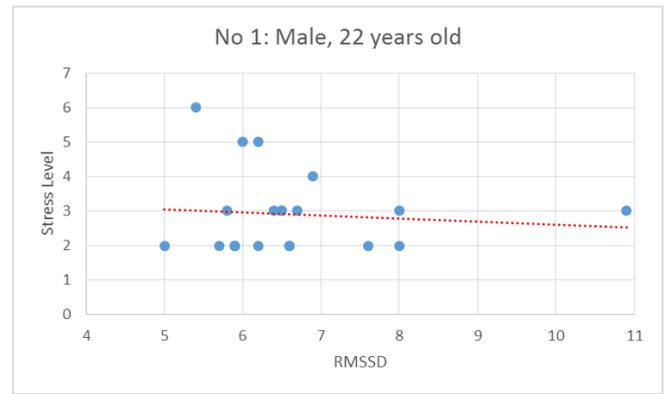

Image 7: RMSSD vs Stress Level #1

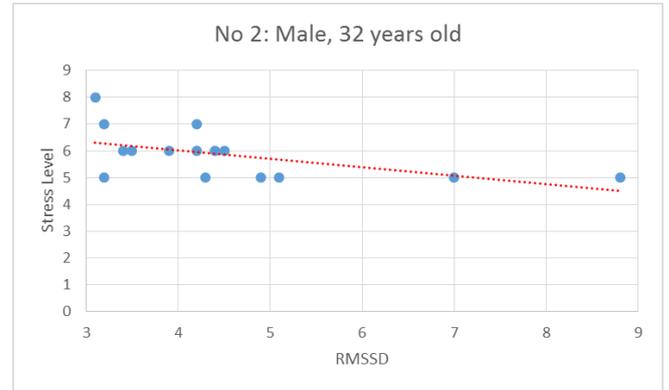

Image 8: RMSSD vs Stress Level #2

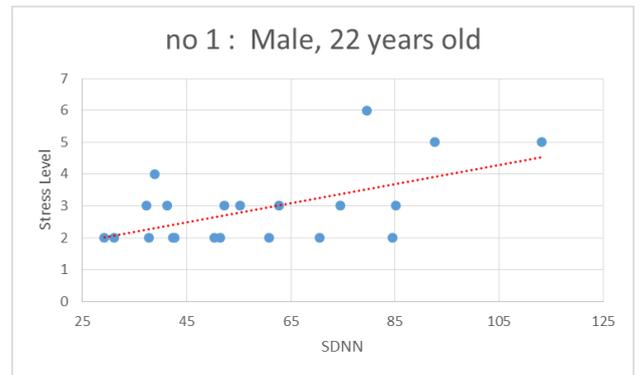

Image 9: SDIBI vs Stress Level #1

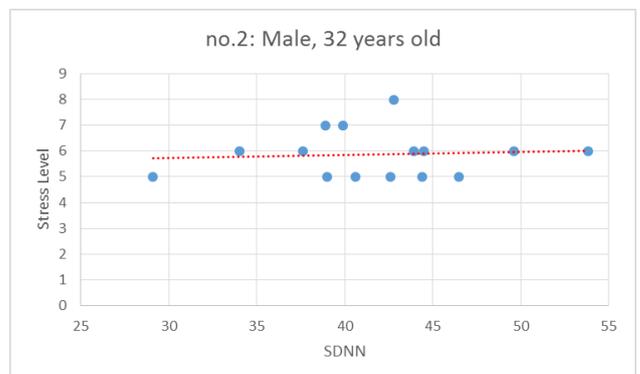

Image 10: SDIBI vs Stress Level #2

And for the influenza:

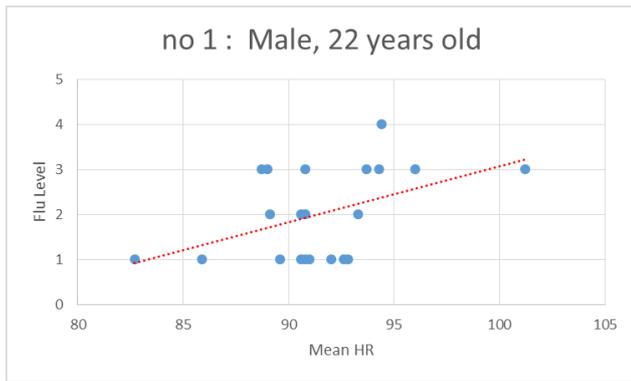

Image 11: Mean HR vs Flu Level #1

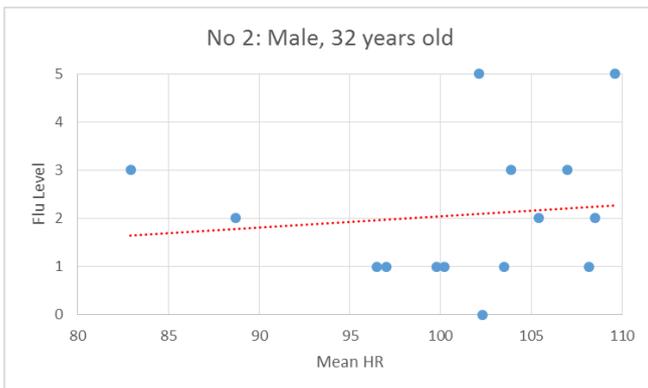

Image 12: Mean HR vs Flu Level #2

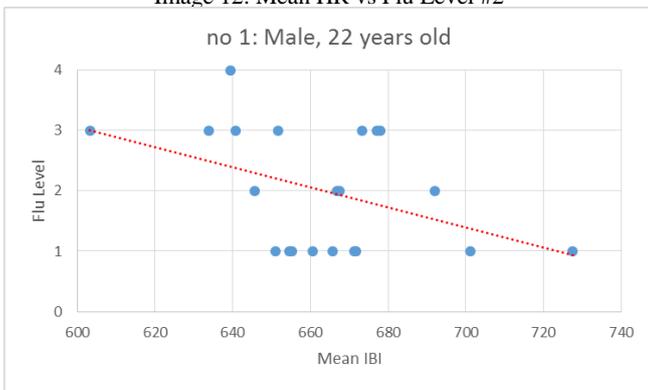

Image 13: Mean IBI vs Flu Level #1

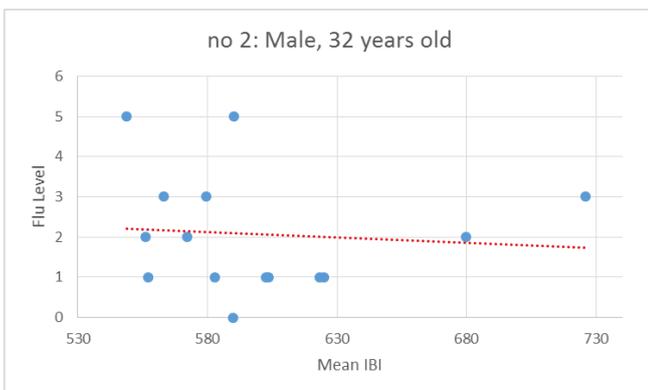

Image 14: Mean IBI vs Flu Level #2

For the classification, it could be very precise if we use the influenza or stress level as a test input. But the problem is, we want to know whether the person is influenza or stress without asking them, so we will use SDevHR and SDevRR as X and Y for stress level. Also, meanHR and MeanIBI assigned as X and Y for flu level. Both data will be used as a learning seed for classification. For the learning output, we will divide the data into 2: sick/stress and normal condition. If the level more than 2, it will considered as stress and if it's 2 or under 2, it will counted as healthy. For the formula, non-stress will be set as -1 and stress will be set as 1. For influenza, if the level more than 1, will be considered as influenza.

## IV. CLASSIFICATION

The classification calculation will be easier than the learning calculation. After we get the bias and weight array on the training, we could calculate the end result. The result will variate depends on the point and the distance to margin. If the result is 0, then the testing point located in the margin line. According to the rule we used when learning, if the test output lead to the positive number, it will classified as influenza/stress and will be classified as healthy if the result is negative. To test the learning curve, we set the input back as test input and then get the precise result:

| Learning Stress Level | Learning Classification | Testing output | Testing Classification | Prediction |
|---|---|---|---|---|
| 3 | 1 | 0.999 | 1 | Correct |
| 2 | -1 | -0.646 | -1 | Correct |
| 5 | 1 | 1.660 | 1 | Correct |
| 5 | 1 | 1.656 | 1 | Correct |
| 6 | 1 | 1.658 | 1 | Correct |
| 4 | 1 | 1.660 | 1 | Correct |
| 2 | -1 | -1.927 | -1 | Correct |
| 3 | 1 | 1.658 | 1 | Correct |
| 3 | 1 | 1.656 | 1 | Correct |
| 2 | -1 | -1.66 | -1 | Correct |
| 2 | -1 | -2.342 | -1 | Correct |
| 2 | -1 | -1.403 | -1 | Correct |
| 3 | 1 | 1.326 | 1 | Correct |
| 2 | -1 | -1.269 | -1 | Correct |
| 2 | -1 | -1.659 | -1 | Correct |
| 2 | -1 | 0.388 | 1 | False |
| 2 | -1 | -2.343 | -1 | Correct |
| 3 | 1 | 0.784 | 1 | Correct |
| 3 | 1 | 0.806 | 1 | Correct |
| 3 | 1 | 1.658 | 1 | Correct |

Table 4: Stress classification result

As we could see, the test could correctly identify 19/20 data and almost predict the false data correctly with score of 0.388. And for the influenza:

| Learning Flu Level | Learning Classification | Testing output | Testing Classification | Prediction |
|---|---|---|---|---|
| 1 | -1 | 1.133 | 1 | False |
| 3 | 1 | 1.629 | 1 | Correct |

| 2 | 1  | 1.740  | 1  | Correct |
|---|----|--------|----|---------|
| 3 | 1  | -0.719 | -1 | False   |
| 3 | 1  | 1.812  | 1  | Correct |
| 1 | -1 | -1.824 | -1 | Correct |
| 1 | -1 | -1.804 | -1 | Correct |
| 3 | 1  | 1.812  | 1  | Correct |
| 2 | 1  | 0.719  | 1  | Correct |
| 1 | -1 | -1.035 | -1 | Correct |
| 2 | 1  | 0.567  | 1  | Correct |
| 2 | 1  | -0.410 | -1 | False   |
| 1 | -1 | -0.719 | -1 | Correct |
| 1 | -1 | -0.999 | -1 | Correct |
| 1 | -1 | -1.516 | -1 | Correct |
| 1 | -1 | -2.188 | -1 | Correct |
| 1 | -1 | -0.410 | -1 | Correct |
| 4 | 1  | 2.277  | 1  | Correct |
| 3 | 1  | 1.812  | 1  | Correct |
| 3 | 1  | 2.268  | 1  | Correct |

Table 5: Influenza classification result

The influenza test predict 17/20 data as correct, less correct data than the stress test. It's due to the influenza no 1 data is converge in the middle and some data crossing such as:

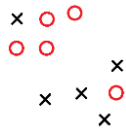

Image 15: Sample of data crossing

And if we graph the problem out to see if it's the problem, with blue and red dot means normal and influenza and 3 dots that mark our false predict:

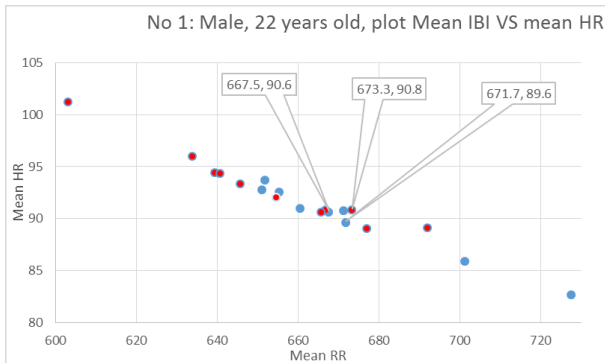

Image 16: Mean IBI vs MeanHR plot

as we can see in the plot that some of the data in the center tend to overlap and quite near. If we zoom the graph we could see that the point is quite near and crossing, making the classification goes wrong.

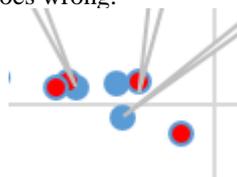

Image 17: Data crossing found

But as the learning data gets more massive, we will predict the data more accurately.

## V. CONCLUSION

Heart rate can be measured from certain place such as chest, arms and neck. Heart rate can be related to the body's condition and mentality. Because HR can be changed very dynamically, data should be taken in a stable condition

Heart rate variance can be used to analyze the human body. HRV includes Time domain, frequency domain, poincare, non-linear and time frequency. For influenza and stress, MeanHR, MeanIBI, SDHR and SDIBI is the best variable from the available recorded data.

Support vector machine can be used to successfully learn and classify data. Lagrangian technique and gaussian kernel used to mapping 2D data into 3D space and get a proper linear classification.

In the end of this paper, heart rate can be used to successfully predict a simple stress or not and influenza or not. Prediction will be more accurate with more data, that's why this result can be used for healthcare application to enhance more people and trends.

REFERENCES

[1] Andrien Ivander Wijaya, Ary S Prihatmanto, Rifki Wijaya. 2016. Shesop Healthcare: Android Application to Monitor Heart Rate Variance, Display Influenza and Stress Condition using Polar H7.

[2] Bernhard Schölkopf, Alexander J. Smola. Learning with Kernels: Support Vector Machines, Regularization, Optimization.

[3] Carlos Guestrin. 2007. SVMs, Duality and the Kernel Trick Machine Learning. Carnegie Mellon University.

[4] European Heart Journal. 1996. Heart rate variability: Standards of measurement, physiological interpretation, and clinical use.

[5] Gary G Berntson. 1997. Heart rate variability: origins, methods and interpretive caveats.

[6] Jason Weston. 2009. Support Vector Machine (and Statistical Learning Theory). NEC Labs America: Princeton USA

[7] Juan F. Ramirez-Villegas, Eric Lam-Espinosa, David F. Ramirez-Moreno. Heart Rate Variability Dynamics for the Prognosis of Cardiovascular Risk

[8] J.T. Ramshur. 2010. HRVAS: Heart Rate Variability Analysis Software University of Memphis.

[9] J.T. Ramshur. 2010. "Design, Evaluation, and Application of HRVAS". University of Memphis.

[10] Markad V. Kamath, Mari A. Watanabe, Adrian R.M. Upton. 2012. Heart Rate Variability (HRV) Signal Analysis: Clinical applications.

[11] Paolo Melillo, Marcello Bracale and Leandro Pecchia. 2011. Nonlinear Heart Rate Variability features for real-life stress detection. Case study: students under stress due to university examination.

[12] R. Berwick. 2003. Guide to Support vector machines (SVMs).

[13] Robert W Levenson, Paul Ekman. 2002. Difficulty does not account for emotion-specific heart rate cahanges in the directed facial action task. Cambridge: USA.

[14] Thorsten Joachims. Learning to Classify Text Using Support Vector Machines.

[15] http://www.fitbit.com/ , 2016

[16] http://www.marcoaltini.com , 2016

[17] http://www.mathworks.com/ , 2016

[18] http://www.heart.org , 2016

[19] http://www.polarusa.com/ , 2016